# Controllable localization and manipulation of optical hollow traps by means of optical-vortex diffraction

Angelsky O.,[1] Bekshaev A.,[1,2] Zenkova C.,[1] Mokhun I.,[1] and Zhang X.[3]

[1]Correlation Optics Department, Chernivtsi National University, Kotsyubinsky str. 2, Chernivtsi 58012, Ukraine
[2]Physics Research Institute, Odesa I.I. Mechnikov National University, V. Zmiienka 2, Odesa 65082, Ukraine, bekshaev@onu.edu.ua
[3]Nankai University, Tianjin, China

**Abstract.** An optical vortex (OV) is coupled with the local intensity zero and is thus a field configuration suitable for "hollow" optical traps and optical tweezers. When an incident circular OV beam experiences diffraction at a rectilinear screen edge (SE), and the SE is far enough from the beam axis, the main consequence is the OV-core displacement from its initial position. Based on the model of incident Laguerre-Gaussian beam, we investigate analytically and numerically the ways of controlling the OV-core position in the diffracted-field by means of changing the SE position with respect to the incident-beam axis. The results show that regulating the SE position with ~1 μm accuracy, controllable OV-core motion with a sub-nanometer accuracy can be realized. Additional control possibilities relating with the incident beam topological charge, wavefront curvature, SE properties (semitransparent screen, phase-step screen) and the distance between the diffraction plane and the observation plane are analyzed.

## 1. Introduction

Optical tools for micro-objects' trapping and manipulation are among the most popular and rapidly developed applications of structured light fields [1,2]. Multiple schemes of optical traps and optical tweezers (OT) are widely used for deliberate capture, prescribed motion, spatial distribution and allocation of various micro- and nanoparticles [1–11]. Sometimes, 3D OTs are realized that perform the "full" 3D localization of a selected micro-object but in many cases, an observed particle is always situated, at least approximately, in a single plane determined by the geometry of the sample (cuvette) [11]. Then, this plane can be considered as the transverse $(x, y)$ plane of a light beam propagating in the longitudinal $z$-direction, and the task of optical manipulation is reduced to the particle's localization within a chosen area of this "target" plane. In such widespread situations, the 2D optical traps and OTs are applicable, which employ the focused paraxial beams with well-developed transverse inhomogeneity.

In such fields, a suspended micro-object experiences the gradient force action depending on its dielectric properties. A dielectric particle with refractive index higher than that of the environment, tends to be localized at the intensity maximum; in contrast, the particles with small refractive index (e.g., air bubbles, oil droplets, biological cells in water), as well as absorbing ones, concentrate at the intensity minima ("hollow traps"). An additional profit of these "hollow OTs" [12–14] is the absence of light-induced damage potentially occurring at the intensity maxima, which is especially important for biological and fragile objects. In such situations, the beams with phase singularities (optical vortices, OVs) [15–18] are especially suitable [1,10]. A characteristic feature of such beams is the existence of a phase singularity ("OV core"). It is a point of the beam cross section $(x, y)$, where the field amplitude possesses an isolated zero (a suitable attribute of the hollow trap). Simultaneously, the optical phase experiences an increment $2\pi m$ on a round trip near the core, $m$ being the integer topological charge of the singularity.

In various OT applications, the crucial issue is the ability of tuning and regulation of the trap position at the target plane with a good thermal, mechanical, and optical stability, as well as suitable and reliable calibration procedure, resulting in high precision and accuracy. Normally, the task of tuning and regulation is fulfilled by a projective optical system including a spatial light modulator (SLM) followed by the high-NA focusing objective [3–10]. There are many specific realizations of such arrangement but all of them obey the similar wave-optics limitations which practically restrict the available accuracy to the values of $\sim 0.1\lambda$ where $\lambda$ is the radiation wavelength [19,20]. In this work, we consider an alternative approach to the spatial tuning of the hollow OT in the target plane, which employs the regularities of the OV beams' diffraction.

Diffraction is one of the most traditional and deeply investigated phenomena of classical optics [19,20]. However, the edge diffraction of circular OV beams [21–34] (see the general scheme in Fig. 1 below) shows many impressive non-trivial details associated with their special physical attributes: helical wavefront shape and transverse energy circulation. In this case, common and well-studied diffraction effects (fringes, transverse diffusion of the light energy, etc.) are supplemented with the OV-specific transformations of the diffracted beam profile. For the purposes of this paper, the most important diffraction effects appear under conditions of weak diffraction perturbation (WDP), when the screen edge (SE) is located far enough from the incident beam axis. Practically, for usual OV beams, the WDP conditions are realized if the screen is distanced from the axis by two or more beam radii measured at the $e^{-1}$ intensity level, and the diffracted beam profile visually preserves the initial circular shape. Nevertheless, the exact OV position shows dramatic transformations, although in microscopic scales:

(i) In a single-charged circular OV, the OV core displaces from the nominal beam axis; the displacement magnitude and direction depend on the SE position and the post-screen propagation distance $z$;

(ii) Like due to any breakdown of the OV-beam circular symmetry [35–38], a high-order OV (with the topological charge $m$, $|m| > 1$) is decomposed into a set of $|m|$ secondary single-charged ones, situated in closely adjacent but different points of the beam cross section, and their positions also evolve with changing the SE position and the post-screen propagation distance;

(iii) At the WDP conditions, when the SE performs a monotonous translation in the transverse direction towards or away from the beam axis [20–22], the perturbed OV cores describe spiral-like trajectories which are very sensitive to the SE position in the transverse plane [29–34].

In particular, the diffraction-induced spiral evolution of the OV cores is one of the most spectacular demonstrations of the rotational properties inherent in the OV beams. On the other hand, the features (i) – (iii) signify that there is a possibility for very fine adjustment of the OV-core position in the diffracted OV beam by means of the rather coarse regulation of the SE location with respect to the incident beam axis. In this paper, we consider the detailed evaluation and numerical characterization of such procedures, keeping in mind the necessities of the hollow optical trapping and the OT techniques.

We start with description of the general scheme for the OV-beam diffraction, quantification of the WDP conditions and presentation of the asymptotic analytical model [29–32] for the diffraction of circular Laguerre-Gaussian (LG) beams [1,15,16] with zero radial index (Section 2). Afterwards, the described theoretical instruments are applied for analysis of the diffracted field and its modifications depending of the SE position with respect to the incident-beam axis; the main attention is paid to the behavior of the OV cores, which form the physical essence of the hollow traps. In this view, the single-charged incident OV beams ($|m| = 1$) appear to be the most promising, and examples of possible OV-core trajectories observable at different diffraction configurations are considered in Section 3. Additional possibilities for the controllable OV-core migration, emerging due to variations of the incident-beam wavefront, are discussed in Section 4. The results are summarized in Conclusion (Section 5), where their general meaning is discussed, as well as possible ways of further practical applications, generalizations and specifications.

## 2. Principal scheme of the OV diffraction

A typical arrangement for observation of the OV diffraction [21–34] is presented in Fig. 1a. As a standard example of the “source” OV beam we consider the circular LG mode [15–18] which can be easily obtained in a laser cavity as well as produced extra-cavity from the circular Gaussian beam. The incident circular OV beam propagates along the axis $z$ (which is simultaneously its symmetry axis) and passes near the diffraction obstacle S situated in the transverse ‘diffraction plane’ $(x_a, y_a)$ orthogonal to the propagation axis. The SE is parallel to the axis $y_a$ and can be translated along the direction $x_a$ so that the axes $(x_a, y_a, z)$ form the Cartesian frame (Fig. 1a). The SE position $a$ with respect to the beam axis and the observation-plane position $z$ with respect to the screen plane are precisely adjustable.

Following to the usual practice reflected in the most of previous works [21–27], we consider the incident monochromatic beam (frequency $\omega$, wavenumber $k = \omega / c$ and wavelength $\lambda = 2\pi/k$, $c$ is the light velocity). Under the paraxial conditions [1,16,18], the beam spatial structure is characterized by the slowly varying complex amplitude $u(x, y, z)$ so that the electric field in any spatial point $(x, y, z)$ at the moment of time $t$ equals to $E = \mathrm{Re}\left[u(x, y, z)\exp(ikz - i\omega t)\right]$. At the diffraction plane ($z = z_a$), the incident beam complex amplitude equals to $u(x_a, y_a, z_a)$, and after passing the screen, experiences the transformation

$$u_a(x_a, y_a) = T(x_a, y_a)u(x_a, y_a, z_a) \tag{1}$$

where $T(x_a, y_a)$ is the spatially inhomogeneous screen transmission coefficient. For the edge diffraction (Fig. 1a), we accept $T(x_a, y_a)$ in the following form:

$$T(x_a, y_a) = \begin{cases} 1, & x_a < a; \\ \tau, & x_a \ge a. \end{cases} \tag{2}$$

In the usual diffraction case, where the screen is opaque and has a sharp edge, $\tau = 0$, but the form (2) includes also the situations of the phase-step diffraction $\tau = \exp(i\varphi)$ [33,34] with arbitrary constant phase shift $\varphi$, or partially transparent screen with $0 < \tau < 1$.

For simplicity, we restrict the consideration to the incident LG beams with zero radial index; in the resulting $LG_{0m}$ modes, the azimuthal index $m$ coincides with the OV topological charge. For them [18,29,30]

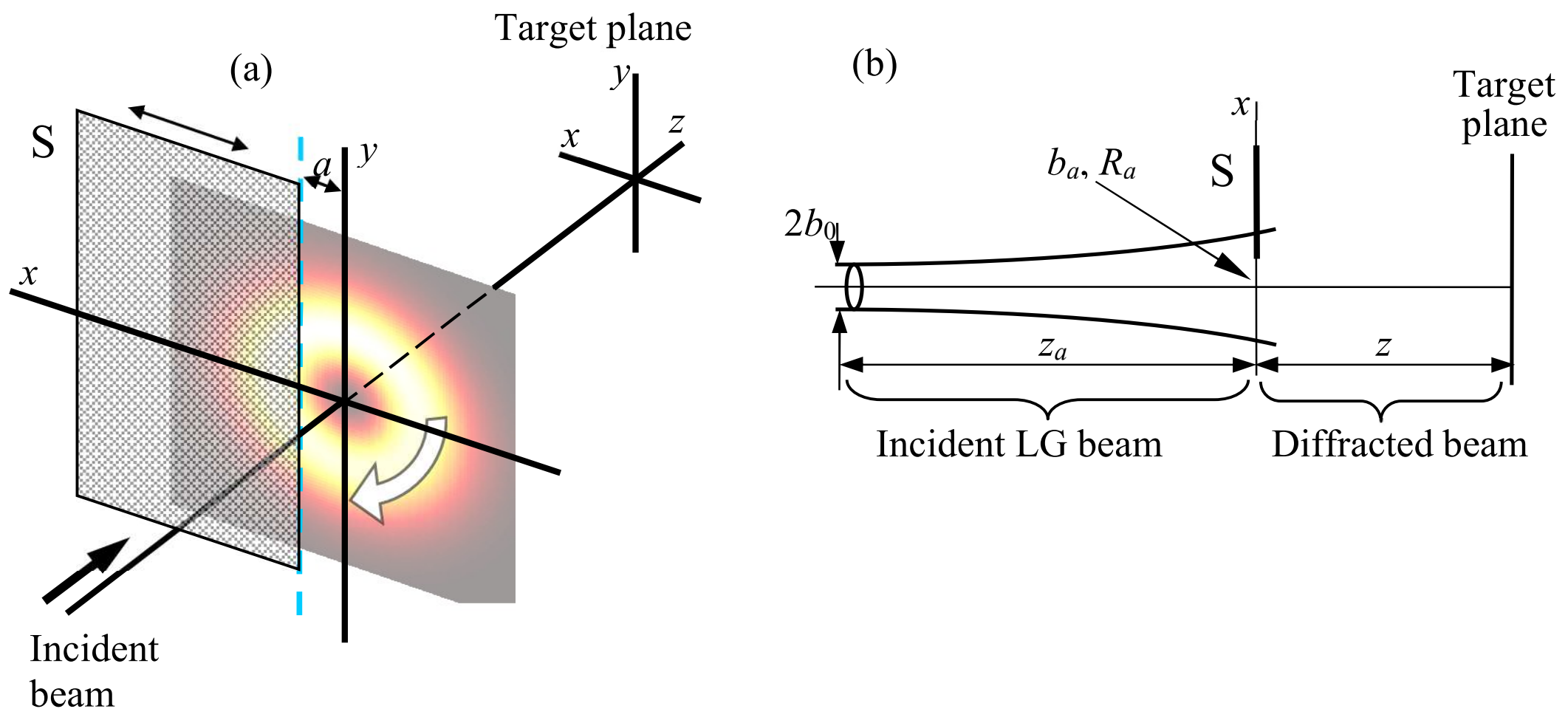

**Fig. 1.** General scheme of the OV beam diffraction: (a) magnified view of the beam screening and the involved coordinate frames, the white arrow shows the transverse energy circulation in a beam with positive topological charge, $m > 0$; (b) scheme of evolution and diffraction of the incident LG beam: S is the diffraction obstacle (the edge is parallel to axis $y$, its adjustable position along axis $x$ is characterized by the distance $a$ from the axis $z$), the diffraction pattern is registered in the observation (target) plane. Further explanations see in text.

$$u(x_a, y_a, z_a) = \frac{(-i)^{|m|+1}}{\sqrt{|m|!}} \left( \frac{z_R}{z_a - iz_R} \right)^{|m|+1} \left( \frac{x_a + i\sigma y_a}{b_0} \right)^{|m|} \exp\left( \frac{ik}{2} \frac{x_a^2 + y_a^2}{z_a - iz_R} \right) \tag{3}$$

where $\sigma = \operatorname{sgn}(m) = \pm 1$, $b_0$ is the Gaussian envelope waist radius, $z_a$ is the distance from the waist cross section to the screen plane (see Fig. 1b), and $z_R = kb_0^2$ is the corresponding Rayleigh length. The helical wavefront is described by the term

$$(x_a + i\sigma y_a)^{|m|} = \sqrt{x_a^2 + y_a^2} \exp\left( im \arctan \frac{y_a}{x_a} \right), \tag{4}$$

the current beam radius $b_a$ and the curvature radius $R_a$ of the spherical wavefront component in the screen plane are determined by the well-known equations [1,15,18]

$$b_a^2 = b_0^2\left(1+\frac{z_a^2}{z_R^2}\right),\quad R_a = z_a + \frac{z_R^2}{z_a}. \tag{5}$$

In the scheme of Fig. 1, the diffracted field at a distance $z$ behind S (target plane in Fig. 1) is described by the Kirchhoff-Fresnel integral [19]

$$u(x,y,z) = \frac{k}{2\pi iz}\int_{-\infty}^{\infty} dy_a \int_{-\infty}^{a} dx_a\, u_a(x_a,y_a)\exp\left\{\frac{ik}{2z}\left[(x-x_a)^2+(y-y_a)^2\right]\right\}. \tag{6}$$

Substitution of Eqs. (1) – (3) into the integral formula (6) gives possibilities for fruitful numerical and analytical investigations of the diffracted field and its relations with the diffraction-scheme parameters. Remarkably, for the WDP conditions, the calculations can be performed based on the asymptotic analytical model valid for $a >> b_a$ [29–32]. Here, we employ this model, and for simplicity (and, as will be shown below, with no limitation of generality), accept the condition

$$z_a = 0, \tag{7a}$$

which implies

$$b_a = b_0,\ R_a = \infty \tag{7b}$$

(the diffraction plane coincides with the waist plane of the incident beam (3)). In this case, the result of integration (6) in the near-axis region $x << b_a$, $y << b_a$ can be presented in the form

$$u(x,y,z) \simeq Br^{|m|}\exp(im\phi) - D(1-\tau)a^{|m|-1}\exp\left[\frac{ika^2}{2}\left(\frac{i}{z_R}+\frac{1}{z}\right)\right] \tag{8}$$

where $r=\sqrt{x^2+y^2}$ and $\phi = \arctan(y/x)$ are the polar coordinates in the target plane,

$$B = \frac{1}{(z-iz_R)^{|m|+1}},\quad D = \sqrt{\frac{i}{2\pi}\frac{k}{z}}(-iz_R)^{-|m|-1}\left[k\left(\frac{1}{z}+\frac{i}{z_R}\right)\right]^{-3/2}. \tag{9}$$

Hence, the expressions for the OV cores' polar coordinates can be easily found via equating (8) to zero:

$$r = \left\{a^{|m|-1}\exp\left(-\frac{a^2}{2b_a^2}\right)\left|\frac{D(1-\tau)}{B}\right|\right\}^{1/|m|}$$
$$= a^{1-1/|m|}\left[|1-\tau|\exp\left(-\frac{a^2}{2b_a^2}\right)\sqrt{\frac{z}{2\pi k}\sqrt{1+\left(\frac{z}{z_R}\right)^2}}\right]^{1/|m|}, \tag{10}$$

$$\phi + M\cos\phi = \frac{1}{m}\left\{\arg\left[D(1-\tau)\right]-\arg B\right\}+\frac{ka^2}{2mz}+\frac{2N}{m}\pi$$
$$= \frac{\arg(1-\tau)}{m}+\frac{1}{m}\left(|m|-\frac{1}{2}\right)\arctan\left(\frac{z}{z_R}\right)+\frac{\pi}{4m}+\frac{ka^2}{2mz}+\frac{2N}{m}\pi,\quad N = 0, 1, \ldots |m|-1 \tag{11}$$

where

$$M = \frac{kra}{mz}. \tag{12}$$

and $N+1$ is the "number" of a secondary single-charged OV formed after diffraction (if $|m| = 1$, there is only one OV corresponding to $N = 0$). The results (10) – (12) are valid in the nearest vicinity of the beam axis $r << b_a$; with this precaution, due to the rapid exponential decay of the field amplitude (3) with growing off-axial distance $a$, they offer an approximately 1% accuracy [29–31] for

$$a \gtrsim 2.5b_a . \tag{13}$$

On the other hand, they obey the conditions for the Fresnel-Kirchhoff approximation, which imply that the distances between the involved points of the diffraction plane and of the target plane should be much higher than the diffraction aperture [19,39]. In application to the LG beam diffraction, these conditions require

$$\left(\frac{z}{b_a}\right)^3 \gg \frac{kb_a}{16\pi} \tag{14}$$

and, for example, if $kb_a \simeq 10^3$, dictate that the distance between the screen plane and the target plane should not be too small, practically $z \gtrsim 10b_a$. The appropriateness of the model (8) – (12) with provisions (13), (14) has been confirmed by direct numerical calculations of the Fresnel-Kirchhoff integral (6) and in OV-diffraction experiments [29–34].

## 3. Diffraction-induced motion of the OV core: diffraction at the beam waist

Equations (9) – (11) enable to reveal all the important details of the diffraction-induced migration of the OV cores observable in the target plane. We illustrate this behavior with the help of several examples of diffracted LG modes (3), satisfying the waist-plane conditions (7); for numerical estimations, we accept the beam parameters

$$k = 10^5 \text{ cm}^{-1}, \quad b_0 = 0.01 \text{ cm}. \tag{15}$$

The accepted value of $k$ approximately corresponds to the He-Ne laser radiation, while the waist radius (~100 μm) corresponds to $(kb_0) = 1000$, which warrants the applicability of the paraxial description [1,16]. The distance $z$ to the target plane is selected within the range 2 cm < $z$ < 10 cm, while the SE position varies near the value $a \approx 2.5b_a$. At these conventions, the requirements (13), (14) are well satisfied, and the formulae (10) – (12) are applicable.

We start with a short comparison of the patterns available at different topological charges of the incident beam. Fig. 2a presents the fragments of the observable OV trajectories for $m$ = 1, 2, and 3 (for beams with opposite signs of $m$, the trajectories can be obtained via the mirror-reflection with respect to the horizontal $x$-axis [29,30,31,32]). The figure shows that the diffraction of an $m$-charged OV, indeed, produces $|m|$ single-charged secondary OVs in the diffracted field. They perform seemingly separate evolutions; however, the intensity "barriers" between their cores are rather low at the considered WDP situation. As a result, the optical traps formed, e.g., by the OV-cores A, B and C, originated from the 3-charged incident OV (green curves), are not well isolated from each other, and rather constitute a single "blurred" trap. In Fig. 2b, the trap boundaries are conventionally characterized by the contours corresponding to the intensity level 5% of the maximum, and the broadening, asymmetry and deformations of the 2nd- and 3rd-order OV "traps" (red and green lines), contrasting from the 1st-order trap (blue contour), are well seen. Of course, such "composite" traps as shown by the red and green lines will be less efficient than the one formed by the "genuine" single-charged OV. Besides, Fig. 2a shows that the same SE increments produce much higher displacements of the secondary OVs (red and green curves) than of the "genuine" single-charged OV (the blue curve inside the dashed square near the coordinate origin). This means that the diffraction of a 1st-order OV is much more suitable for fine regulation of the OV-core position via the SE translation. Accordingly, in further consideration we focus on the case $|m| = 1$, and analyze the situation inside the target plane fragment denoted by the dashed square of Fig. 2a.

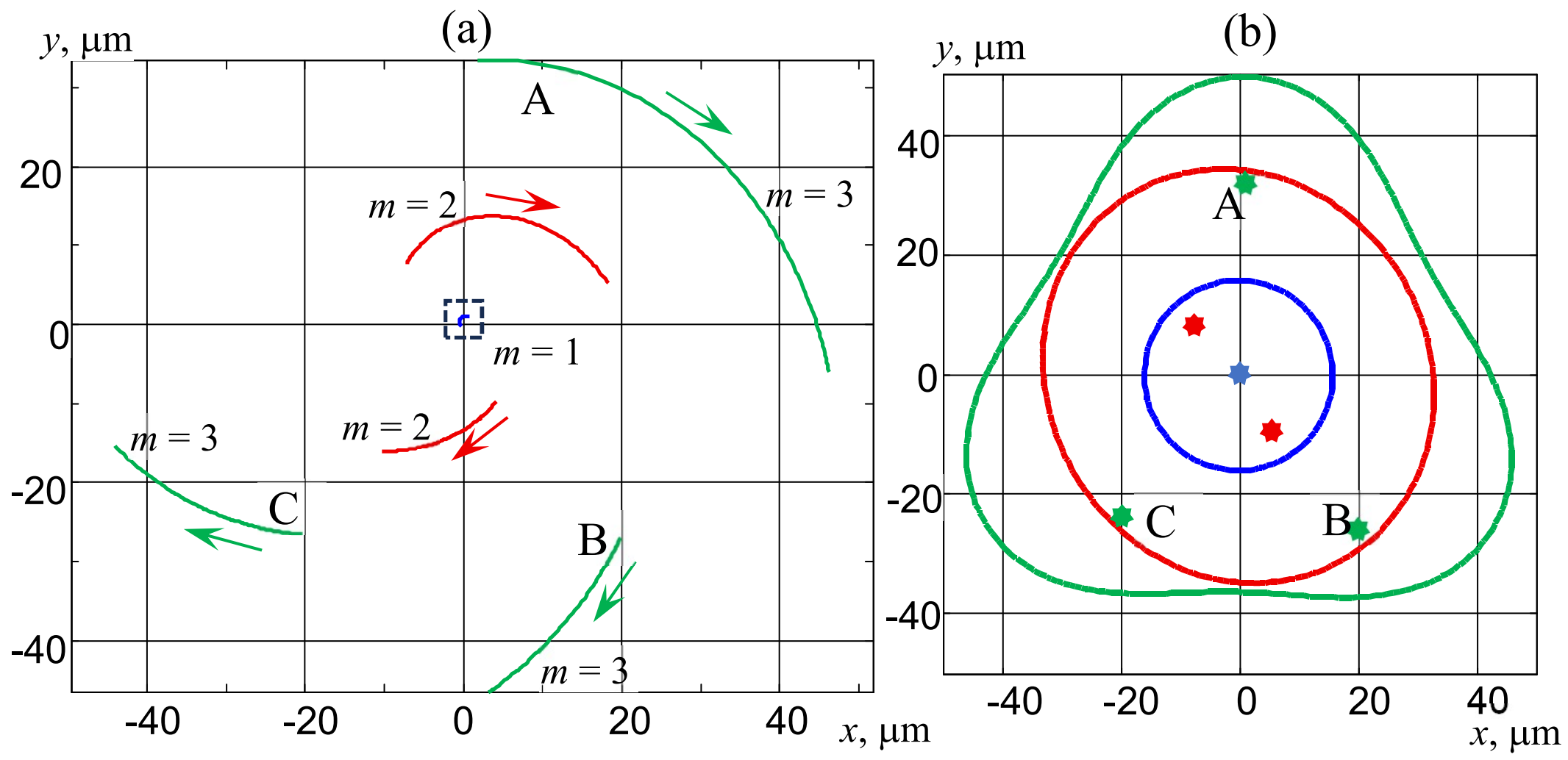

**Fig. 2.** Characteristics of the OV-core positions and the near-core intensity distributions in the target plane $z = 5$ cm (see Fig. 1) for diffraction at the opaque-screen ($\tau = 0$ in Eqs. (10), (11)) of an LG beam (3) with the topological charge: $m = 1$ (blue); $m = 2$ (red); $m = 3$ (green). (a) OV-core trajectories observable when SE translates from $a = 3b_a$ to $a = 2.5b_a$ (the trajectory for $m = 1$ is shown by the small blue curve inside the dashed square); arrows show the directions of the OV migration. (b) Constant-intensity contours corresponding to 5% of the maximum (calculated from (8), (9)) and the approximate OV-core positions (stars) for $a = 3b_a$ (starting point of the evolution illustrated in (a)). Other explanations see in the text.

Further details relating the diffracted field of a single-charged incident LG beam (3) are presented in Fig. 3 where additional characteristics of the diffraction-induced OV-core displacements are illustrated. As has been established earlier [29–34], the SE translation induces a spiral-like motion of the OV core in the $(x, y)$-plane; with growing $z$, the spiral curvature decreases, and the trajectory 'as a whole' shows a higher deviation from the beam axis (Fig. 3a). Numerical evaluation of the interrelations between the OV position and the SE displacement $a$ can be obtained with the help of Fig. 3b where the $a$-dependent behavior of the OV-core Cartesian coordinates ($x = r\cos\phi$, $y = r\sin\phi$) is shown (solid lines). Additionally, the derivatives $dx/da$, $dy/da$ are presented by the dashed curves, which illustrate the sensitivity of the OV-core location to the screen position $a$ (see Fig. 1). According to these data, the SE shift $\Delta a \approx 0.01b_a = 1$ μm results in the OV-core shift $(|\Delta x|, |\Delta y|) \approx 0.02$ μm. This fact illustrates the general useful regularity: Controlling the SE position with an accuracy $\delta$ enables to control the OV localization with the accuracy $\sim 0.02\delta$, and this is obtained from the first occasional example.

Undoubtedly, special efforts aimed at the accuracy enhancement can essentially improve this evaluation. The simplest ways may be associated with utilization of a semitransparent screen: the data of Fig. 3 are obtained for $\tau = 0$ (see Eq. (2)) whereas, in agreement with Eq. (10) for $|m| = 1$, $r$ is proportional to $(1 - \tau) < 1$, and the absolute values of the controllable OV-core displacement $(|\Delta x|, |\Delta y|)$ can be easily reduced several times for the same $\Delta a$, if a screen with $0.5 < \tau < 1$ is used.

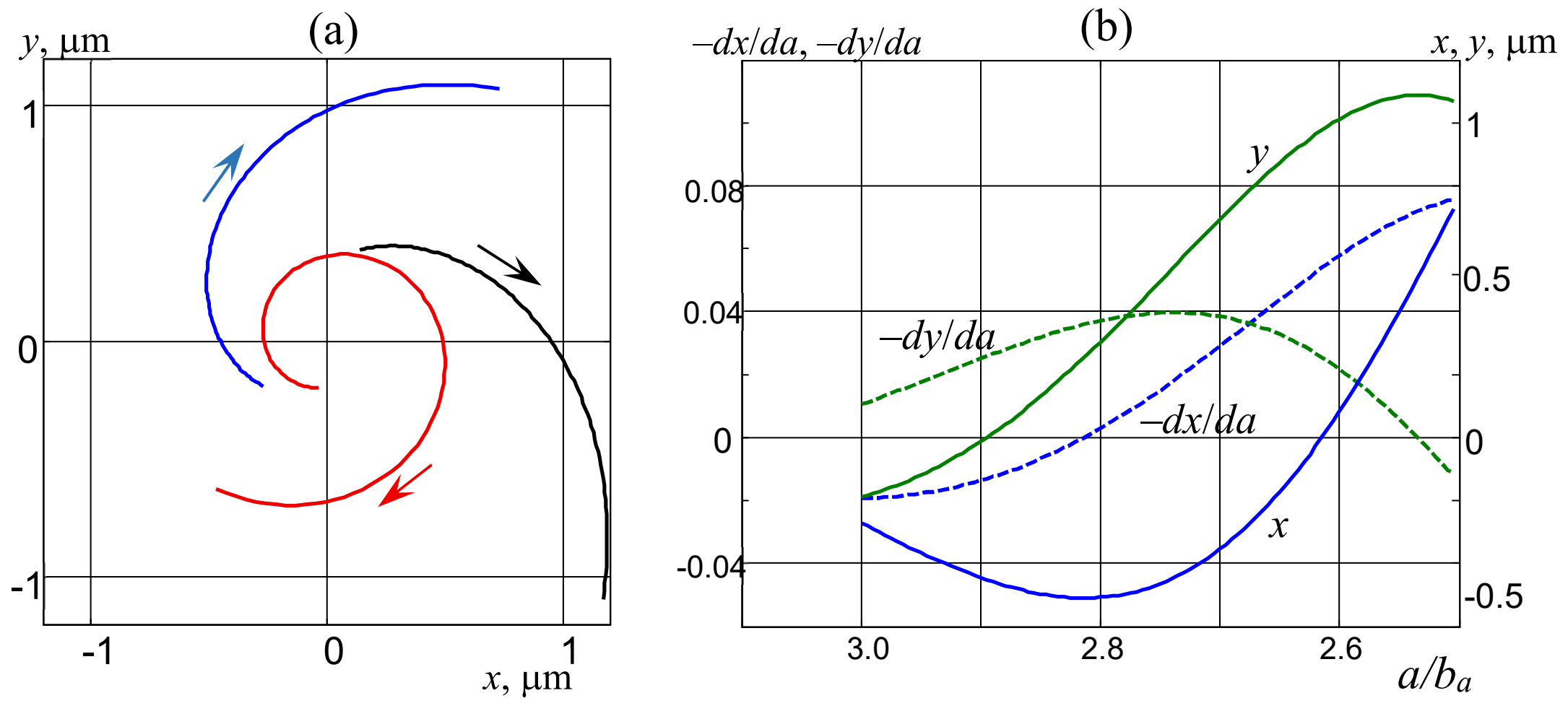

**Fig. 3.** Characteristics of the diffracted-field OV-core motion for the incident $LG_{01}$ beam (3) with the topological charge $m$ = 1 and parameters (7a), (7b), (15), observable when the opaque-screen ($\tau$ = 0 in Eqs. (10), (11)) translates from $a = 3b_a$ to $a = 2.5b_a$: (a) OV-core trajectory in the target plane $z$ = 2 cm (red line), $z$ = 5 cm (blue line) – the same as the blue curve inside the dashed square at the center of Fig. 2a, and $z$ = 7 cm (black line); arrows show the directions of the OV motion when the SE moves from the beam periphery towards the axis; (b) $a$-dependent evolution of the OV-core Cartesian coordinates $x$, $y$ (solid lines, right scale) and of the derivatives $-dx/da$, $-dy/da$ (dashed lines, left scale) in the target plane $z$ = 5 cm. Negative signs of the derivatives reflect that the observed evolution is caused by the SE translation from higher to lower $a$ (note the reverse horizontal scale in (b)).

However, high sensitivity of the OV trajectory to the SE position is not the only characteristic important for the OT controllable localization. Fig. 3a testifies that the simple SE translation induces rather intricate curvilinear OV trajectories whose behavior strongly depends on the post-screen propagation distance $z$. Generally, this is not a serious defect for possible OV tuning because, in cases of high accuracy, the expected OV-core deviations from the assigned point are so small that the trajectory can be replaced by the rectilinear segment whose direction is dictated by the local values of $dx/da$, $dy/da$ at this point. Moreover, the dependence of the OV trajectories on $z$ can be used for deliberate assignment of the direction in which the OV moves in response to certain standard SE displacement. Nevertheless, the search of conditions where the OV core performs a less intricate, more predictable, preferably straight-line trajectories, looks a reasonable task. This can be implicated, e.g., by employing an additional "degree of freedom", which appears if we remove the restrictions (7) and admit the incident LG beams with spherical wavefronts [30].

## 4. Description of the OV displacements for the incident LG beam with spherical wavefront

The above consideration was based on the simplifying assumptions (7), which meant that the incident-beam wavefront curvature is exclusively associated with the helical phase (4) and has no spherical component; upon the relevant discussion, it was emphasized that the conditions (7) imply no limitation of generality. Now we elucidate this argument and employ it for analysis of the influence which can be caused by the spherical wavefront curvature. To this end, we involve another simple rule for transformation of the diffracted beam complex amplitude that occurs if the incident beam is modified according the equation

$$u(x_a, y_a) \to u(x_a, y_a) \exp\left( ik \frac{x_a^2 + y_a^2}{2R_a} \right) \qquad (16)$$

(i.e., the spherical wavefront component is added to the beam preserving the same intensity profile [28]). Practically, the transformation (16) can be realized by a proper spherical lens placed immediately before or behind the diffraction plane. In such a situation, if the initial distribution $u(x_a, y_a)$ ("prototype beam") produces the diffracted field with complex amplitude $u_p(x,y,z)$, the modified initial beam (16) produces the diffracted field described by

$$u(x,y,z) = \frac{1}{1+\dfrac{z}{R_a}} \exp\left[ik\frac{x^2+y^2}{2(z+R_a)}\right] u_p\left(\frac{x}{1+z/R_a}, \frac{y}{1+z/R_a}, \frac{z}{1+z/R_a}\right) \tag{17}$$

(this can be shown immediately from the Fresnel-Kirchhoff integral (3) [28,30]). To elucidate the meaning of this rule, let us suppose that the incident prototype beam $u(x_a, y_a)$, being diffracted, produces in the cross section $z_0$ behind the screen such complex amplitude distribution, for which the OV core is situated in the point ($x = x_0$, $y = y_0$). Hence, Eq. (17) dictates that the modified incident beam (16) produces the diffracted field in whose cross section

$$z = \frac{z_0}{1 - z_0/R_a} \tag{18}$$

the OV core is located in the point

$$x = \frac{x_0}{1 - z_0/R_a}, \quad y = \frac{y_0}{1 - z_0/R_a}. \tag{19}$$

In application to equations (9) – (11) this rule yields

$$r = \left[a\left(1+\frac{z}{R_a}\right)\right]^{1-1/|m|} \left[|1-\tau| \exp\left(-\frac{a^2}{2b_a^2}\right)\sqrt{\frac{z}{2\pi k}\sqrt{\left(1+\frac{z}{R_a}\right)^2+\left(\frac{z}{z_R}\right)^2}}\right]^{1/|m|}, \tag{20}$$

$$\phi + M\cos\phi = \frac{\arg(1-\tau)}{m} + \frac{1}{m}\left(|m| - \frac{1}{2}\right)\arctan\left[\frac{z}{z_R(1+z/R_a)}\right] + \frac{\pi}{4m} + \frac{ka^2}{2mz} + \frac{ka^2}{2mR_a} + \frac{2N}{m}\pi . \tag{21}$$

In particular, the far field of the diffracted prototype beam is realized in the modified-beam diffraction at

$$z = -R_a . \tag{22}$$

This distance between the diffraction plane and target plane can be physically realized if $R_a < 0$ (converging wavefront). Actually, exactly such a situation occurs if a lens of the focal length $f_a = -R_a$ is placed in the diffraction plane, and the target plane is assigned at the focal plane of the lens: there, the Fraunhofer pattern of the diffracted field is reproduced.

This situation is illustrated in Fig. 4a that shows several possible OV-core trajectories achievable at the target plane (22), upon diffraction of a single-charge LG mode (3) subjected to transformation (16) with $R_a = -z$. At the standard conditions of an opaque screen ($\tau = 0$ in Eqs. (20), (21)), the spiral-like trajectory of Fig. 3a degenerates into the straight line parallel to the SE (blue lines); in the geometry of Fig. 1, the SE translation from the beam periphery to the axis induces the OV-core motion along the $y$-axis (at $m = 1$), and opposite to it (at $m = -1$). This rectilinear motion is performed almost uniformly, slightly accelerating with decreasing $a$, as can be seen from the curves of Fig. 4b. Available value of $|dy/da| \approx 0.01$ is even lower than in the case of Fig. 3, enabling the accuracy of the OV-core localization $0.01\delta$ if the SE position is tuned with accuracy $\delta$. Of course, the possibilities of further improving the accuracy by using a semitransparent screen, discussed in Section 3 and associated with the term $(1 - \tau)$ in Eqs. (20), (21), are equally applicable in this case.

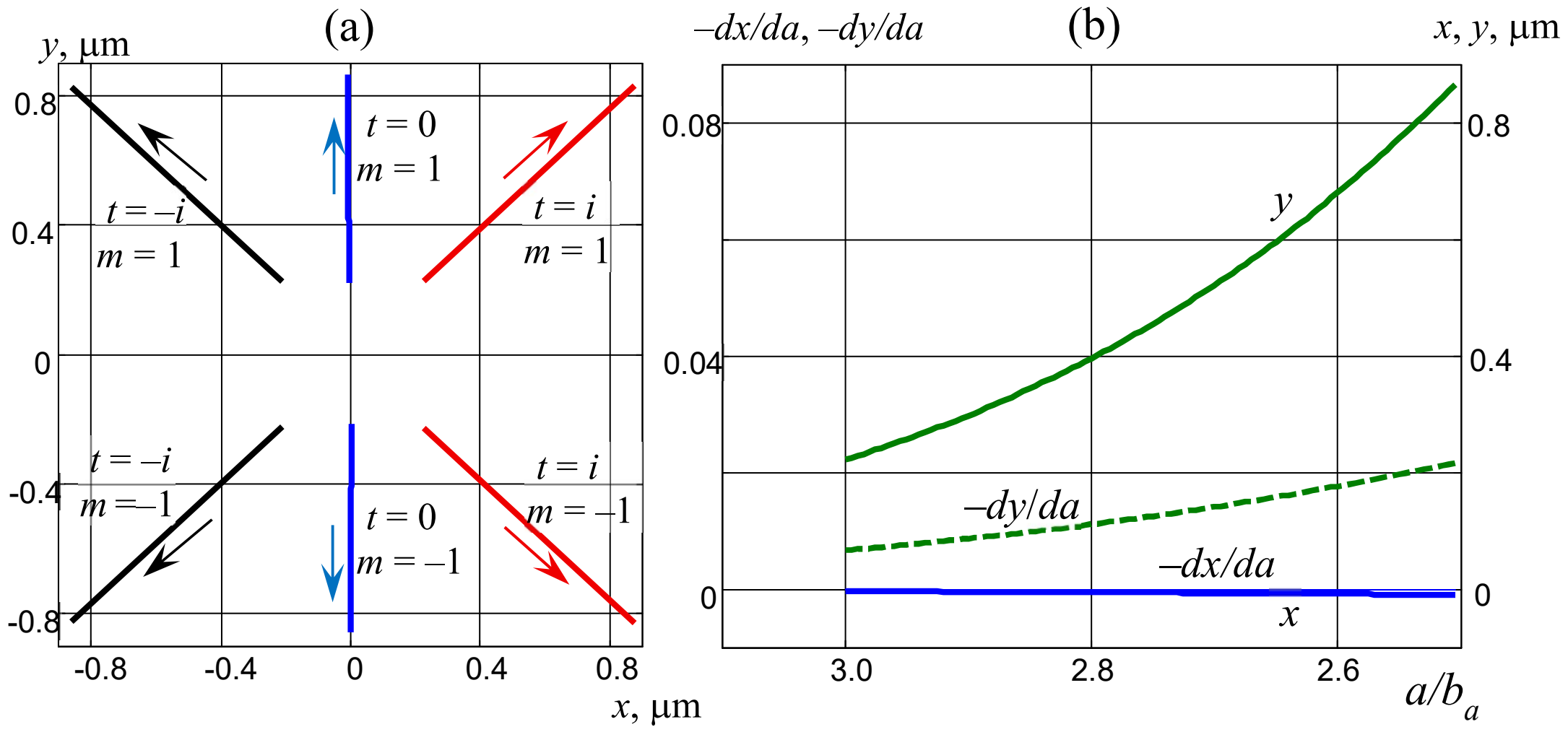

**Fig. 4.** Characteristics of the OV-core behavior in the diffracted field for the incident LG beam (3) with parameters (7a), (7b), (15), transformed according to Eq. (16) with $R_a = -5$ cm, observable in the target plane $z = 5$ cm while the SE translates from $a = 3b_a$ to $a = 2.5b_a$ (see Fig. 1): (a) OV-core trajectory in the target plane for the screen with $\tau = 0$ (blue), $\tau = \exp(i\pi/2)$ (red), $\tau = \exp(-i\pi/2)$ (black) and for positive ($m = 1$) and negative ($m = -1$) topological charge; arrows show the directions of the OV motion when the SE moves from the beam periphery towards the axis; (b) $a$-dependent evolution of the OV-core Cartesian coordinates $x$, $y$ (solid lines, right scale) and of the derivatives $-dx/da$, $-dy/da$ (dashed lines, left scale) for $\tau = 0$ and $m = 1$ (blue graphs for $x$ and $dx/da$ practically do not differ from the horizontal zero-line). Other conventions are as in Fig. 3b.

Remarkably, not only the vertical rectilinear OV trajectories are available in the target plane (22). Additional possibilities are coupled with the diffraction at a rectilinear phase step [33,34] realized in the scheme of Fig. 1a if $\tau = \exp(i\varphi)$ in Eq. (2). For example, at $\varphi = \pi/2$, the OV position evolves along the bisectors of the 1st and 4th quadrants, regarding the OV sign, $m = \pm 1$ (red lines in Fig. 4a); at $\varphi = -\pi/2$, the OV moves along the bisectors of the 2nd and 3rd quadrants (black lines). For other phase-step values $\varphi$, intermediate OV-core trajectories can be obtained.

## 5. Conclusion

The examples and calculations considered in the above Sections support the idea that the predictable and controllable adjustment of the OV-core location can be realized in the process of edge diffraction of circular OV-beams (3) under the WDP conditions (13). Beside the controllable OV-core localization, the WDP conditions offer the specific advantage of approximately preserving the initial circular symmetry of the near-core intensity distribution [29–32] (see, for example, the blue equal-intensity contour in Fig. 2b), thus providing appropriate conditions for the usual trapping and manipulating OT functions. The results of this paper illustrate the main principles and testify for their general validity. At the same time, the specific details of corresponding arrangements and procedures may be modified in agreement with the special tasks and necessities.

The above-described examples represent some simplest and most evident realizations of the proposed ideas. However, the presented mathematical model, resulting in the main relations (10), (11) and (20), (21), offers a wide range of practical possibilities associated with the proper adjustment of the incident-beam configuration ($b_a$, $R_a$) and the diffraction parameters ($\tau$, $a$, $z$). Upon proper selection of the beam structure and the diffraction geometry, a lot of specific OV-core trajectories in the diffracted field can be realized, which may be useful for the OT assembly and operation, as well as for specific solutions for the micro-object allocation and manipulation. Particularly, the immediate and easily realizable ways of further development can be based on involving the "complex" semitransparent screens ($\tau = |\tau|\exp(i\varphi)$) with properly adjusted transparency $0 < |\tau| < 1$ and phase-step value $\varphi$.

The results presented in Figs. 3, 4 show that regulation of the SE position with accuracy $\delta$ enables to control the OV-core position in the diffracted field (and, consequently, manipulate the corresponding hollow-trap position) with the accuracy $\delta/\gamma$ where the "accuracy-enhancement coefficient" $\gamma \approx 10^2$. Moreover, the simple considerations involving the semitransparent screen (Section 3) indicate that this value can be easily improved by the order of magnitude, promising $\gamma \approx 10^3$. This means that, for example, the SE positioning with 1 μm accuracy, which is quite achievable with existing mechanical driving tools [3–10], produces an impressive OT-localization with accuracy ~1 nm. But this is not the final limit. If necessary (and this is the usual situation), the proposed diffraction scheme of Fig. 1 can certainly be supplemented by an optical projection system which forms the target-plane image with appropriate reduction of the transverse geometric size and proportional enhancement of the OT localization accuracy to the sub-nanometer scale. Additional prospects may be associated with employment of more "gentle" OV-diffraction schemes, e.g., diffraction at a "soft" edge [40].

In conclusion, we would like to note that the model of incident LG beam (3), accepted in this paper, puts no principal restrictions, and the similar ideas can be applied to other types of OV beams, e.g., Bessel beams, Kummer beams, etc. The Bessel beams are widely applicable in the OT techniques [13,14]; the Kummer-type beams, or "Hypergeometric modes" are produced by such a widespread tools for the OV-beam generation as the "fork" holograms or spiral phase plates [41–47]. A characteristic feature of such beams is the slower intensity fall-off at the beam periphery: the intensity decays according to a power law instead of exponentially in (3), and the decay is non-monotonous but of oscillatory character. As a consequence, the WDP conditions occur, generally, at higher $a$ than is dictated by the exponential model (Eq. (13)); the derivatives of the diffraction-induced OV-core displacement $|dx/da|$, $|dy/da|$ are, generally, higher than presented and discussed in Figs. 3, 4, and the corresponding OV-core trajectories expose small oscillations and seeming irregularities [29,31,32,34]. At first glance, these peculiarities are unfavorable for the OT applications; on the other hand, they open a lot of new possibilities which may find practical implementations. If necessary, these and other possible generalizations (for example, those related to the asymmetry of the incident OV beams [36,37]) can be performed on the base of the approach formulated and developed in this paper.

**Funding.** This research was supported by the National Research Foundation of Ukraine (NRFU), project No. 2025.06/0086 "Innovative Methods and Systems of Laser Communication in Turbulent Atmospheric Environments Based on Phase Singularities”.

**Acknowledgments.** The authors are grateful to the Department of Correlation Optics at Yuriy Fed'kovych Chernivtsi National University for encouraging support of this work.

**Disclosures.** The authors declare no conflicts of interest.